\documentclass[sigconf,authorversion]{acmart}

\AtBeginDocument{%
  }

\copyrightyear{2025}
\acmYear{2025}

\setcopyright{acmlicensed}
\acmConference[ASIA CCS '25]{ACM Asia Conference on Computer and Communications Security}{August 25--29, 2025}{Hanoi, Vietnam}
\acmBooktitle{ACM Asia Conference on Computer and Communications Security (ASIA CCS '25), August 25--29, 2025, Hanoi, Vietnam}
\acmDOI{10.1145/3708821.3733869}
\acmISBN{979-8-4007-1410-8/2025/08}

\usepackage{algorithm}
\PassOptionsToPackage{full}{textcomp}
\definecolor{indigo}{RGB}{51,34,136}
\definecolor{cyan}{RGB}{136,204,238}
\definecolor{teal}{RGB}{68,170,153}
\definecolor{green}{RGB}{17,119,51}
\definecolor{olive}{RGB}{153,153,51}
\definecolor{sand}{RGB}{221,204,119}
\definecolor{rose}{RGB}{204,102,119}
\definecolor{wine}{RGB}{136,34,85}
\definecolor{purple}{RGB}{170,68,153}
\definecolor{palegrey}{RGB}{221,221,221}
\colorlet{plotcolor0}{indigo}
\colorlet{plotcolor1}{cyan}
\colorlet{plotcolor2}{teal}
\colorlet{plotcolor3}{green}
\colorlet{plotcolor4}{olive}
\colorlet{plotcolor5}{sand}
\colorlet{plotcolor6}{rose}
\colorlet{plotcolor7}{wine}
\colorlet{plotcolor8}{purple}
\colorlet{plotcolorbad}{palegrey}

\usepackage{listings}

\definecolor{spec0}{rgb}{0, 0.4470, 0.7410}
\definecolor{spec1}{rgb}{0.8500, 0.3250, 0.0980} 	          	
\definecolor{spec2}{rgb}{0.9290, 0.6940, 0.1250} 	
\definecolor{spec3}{rgb}{0.4940, 0.1840, 0.5560}          	
\definecolor{spec4}{rgb}{0.4660, 0.6740, 0.1880} 	          
\definecolor{spec5}{rgb}{0.3010, 0.7450, 0.9330} 	        
\definecolor{spec6}{rgb}{0.6350, 0.0780, 0.1840}

\definecolor{codegreen}{rgb}{0,0.6,0}
\definecolor{codegray}{rgb}{0.5,0.5,0.5}
\definecolor{codepurple}{rgb}{0.58,0,0.82}
\definecolor{backcolour}{rgb}{0.95,0.95,0.92}
\definecolor{onespin}{RGB}{95,0,142}

\usepackage{rotating}
\newcommand{\refsec}[1]{Sec.~\ref{sec:#1}}
\newcommand{\reffig}[1]{Fig.~\ref{fig:#1}}

\usepackage{subcaption}

\usepackage{multirow}

\usepackage{makecell}

\usepackage{stmaryrd}

\begin{document}
%
%
\title{Okapi: Efficiently Safeguarding Speculative Data Accesses in Sandboxed Environments}

\author[P. Schmitz]{Philipp Schmitz}
\orcid{0009-0008-5342-3847}
\affiliation{
    \institution{RPTU Kaiserslautern-Landau}
    \city{Kaiserslautern}
    \country{Germany}
}
\email{philipp.schmitz@rptu.de}

\author[T. Jauch]{Tobias Jauch}
\orcid{0009-0009-6547-1544}
\affiliation{
    \institution{RPTU Kaiserslautern-Landau}
    \city{Kaiserslautern}
    \country{Germany}
}

\author[A. Wezel]{Alex Wezel}
\orcid{0000-0002-3462-0094}
\affiliation{
    \institution{RPTU Kaiserslautern-Landau}
    \city{Kaiserslautern}
    \country{Germany}
}

\author[M. R. Fadiheh]{Mohammad R. Fadiheh}
\orcid{0000-0003-0214-2486}
\affiliation{
    \institution{Stanford University}
    \city{Stanford}
    \state{CA}
    \country{USA}
}

\author[T. Tiemann]{Thore Tiemann}
\orcid{0000-0001-9018-4226}
\affiliation{
    \institution{University of Lübeck}
    \city{Lübeck}
    \country{Germany}
}

\author[J. Heller]{Jonah Heller}
\orcid{0009-0004-9987-2403}
\affiliation{
    \institution{University of Lübeck}
    \city{Lübeck}
    \country{Germany}
}

\author[T. Eisenbarth]{Thomas Eisenbarth}
\orcid{0000-0003-1116-6973}
\affiliation{
    \institution{University of Lübeck}
    \city{Lübeck}
    \country{Germany}
}

\author[D. Stoffel]{Dominik Stoffel}
\orcid{0000-0002-8180-9738}
\affiliation{
    \institution{RPTU Kaiserslautern-Landau}
    \city{Kaiserslautern}
    \country{Germany}
}

\author[W. Kunz]{Wolfgang Kunz}
\orcid{0000-0002-6612-2946}
\affiliation{
    \institution{RPTU Kaiserslautern-Landau}
    \city{Kaiserslautern}
    \country{Germany}
}

%
\begin{abstract}

  This paper introduces Okapi, a new hardware/software cross-layer
  architecture designed to mitigate Transient Execution Side Channel
  (TES) attacks, including Spectre variants, in modern computing
  systems. %
  Okapi provides a hardware basis for secure speculation in sandboxed environments and can replace expensive speculation barriers in software.
  
  At its core, Okapi allows for speculative data accesses to a memory
  page only after the page has been accessed non-speculatively at least once
  by the current trust domain. %
  The granularity of the trust domains can be controlled in software to achieve different security
  and performance trade-offs. %
  For environments with less stringent security needs, Okapi's
  features can be deactivated to remove all performance overhead. %
  
  Without relying on any software modification, the Okapi hardware
  features already provide full protection against TES breakout
  attacks, e.g., by Spectre-PHT or Spectre-BTB, at a thread-level
  granularity. %
  This incurs an average performance overhead of only 3.17\,\% for the
  SPEC~CPU2017 benchmark suite. %
  
  Okapi introduces the \emph{OkapiReset} instruction for additional
  soft-ware-level security support. %
  This instruction allows for fine-grained sandboxing with custom
  program sizes smaller than a thread, resulting in 2.34\,\% performance overhead in our WebAssembly runtime experiment. %
  \par
  On top, Okapi provides the possibility to eliminate poisoning
  attacks. %
  For the highest level of security, the \emph{OkapiLoad} instruction
  prevents confidential data from being added to the trust domain
  after a sequential access, thereby enforcing \textit{weak
    speculative non-interference}. %
  In addition, we present a hardware extension that limits the
  exploitable code space for Spectre gadgets to well-defined sections
  of the program. %
  Therefore, by ensuring the absence of gadgets in these sections,
  developers can tailor their software towards achieving beneficial
  trade-offs between the size of a trust domain and performance. %
  \end{abstract}
  
\maketitle

%


\section{Introduction}
\label{sec:introduction}

Transient Execution Side Channel (TES) attacks, including numerous variants
of Spectre~\cite{2019-KocherHorn.etal}, persist as serious threats in
modern computing systems. %
These vulnerabilities enable attackers to exploit side effects of
speculative execution by tricking the processor into 
revealing information that is not accessed by the sequential semantics of the
program. %
Since TES attacks exploit high-end features of modern
microarchitectures such as speculative and out-of-order execution,
eliminating them without introducing a significant performance penalty continues to be a major challenge. %
The most dangerous Spectre attacks involve so-called \textit{universal read gadgets} that are capable of revealing the contents of any address in the entire memory space. %
\par
A widely used software mechanism to restrict a program's accessible resources is \textit{sandboxing}~\cite{1993-WahbeLucco.etal}. %
Untrusted code is often recommended to be executed within a sandbox~\cite{2024-google-sandbox} in order to prevent it from stealing data or compromising the system~\cite{2019-nytimes}. %
As long as no additional measures are in place, Spectre-style attacks pose a severe threat to such software-level isolation techniques~\cite{2022-SchwarzlBorello.etal}. %
Attackers can either perform a \textit{breakout attack}, i.\,e., bypassing sandbox boundaries to access confidential data, or a \textit{poisoning attack}, i.\,e., tricking a victim sandbox into transiently leaking its own memory content~\cite{2021-NarayanDisselkoen.etal}. %
These attacks are particularly relevant to runtime environments that execute multiple (distrusting) tenants within a single thread. %
Securing runtime environments efficiently against breakout attacks is a major concern in industrial applications, e.\,g., serverless computing~\cite{2022-SchwarzlBorello.etal, 2024-cloudflare-security}. %
\par
Although microarchitectural patches have found adoption in commercial
CPUs to mitigate Meltdown and MDS attacks, Spectre attacks are still
primarily addressed by countermeasures at the operating-system and
compiler levels. %
These mitigations entail a high performance
overhead~\cite{2018-Turner, 2019-AmitJacobs.etal} but are used to
patch legacy systems nonetheless. %
This is primarily because secure speculation techniques in hardware,
to date, necessitate significant modifications to the
microarchitecture and the memory subsystem (additionally incurring
substantial performance overheads) which impede their adoption in
industrial settings. %

\subsection{Contribution}
\label{sec:contribution}
The main contributions of the paper are summarized as
follows: 
\begin{itemize}
\item We introduce Okapi, a new hardware/software cross-layer architecture that efficiently protects a system against all breakout attacks at a chosen granularity. With only 0.6\,\% additional cell area, Okapi achieves an exceptionally low performance overhead of only 2.34\,\% for securing a WebAssembly runtime.
\item Okapi also provides full protection against poisoning attacks, if certain assumptions for the software are fulfilled. In \refsec{poisoning} we describe software measures to fulfill these assumptions and in \refsec{sweet-spots} we discuss possible trade-offs between different levels of security and the corresponding overheads in hardware and performance.
\item In combination, the Okapi hardware and software measures are capable of mitigating TES attacks, ranging from a low-overhead countermeasure against Spectre-PHT up to providing  \emph{weak speculative non-interference}~\cite{2021-GuarnieriKoepf.etal}. 
\end{itemize}

At its core, the Okapi architecture extends the Page Table
Entries 
in the Translation Look-aside Buffer~(TLB) with an
additional bit, the \textit{safe access bit}, that marks a data page
whenever one of its addresses is translated legally, i.\,e.,
non-speculatively, without causing an exception, in the currently
running thread. %
If this bit is set, the CPU is allowed to speculatively load from
addresses in this page. %
Conversely, Okapi blocks speculative execution of loads from
pages that do not have the \emph{safe access bit} set and thereby
prevents formation of \emph{any} transient execution side channel
based on the content of these pages. %
Whenever there is a switch between software threads, all \textit{safe
  access bits} in the TLB are reset to prevent the misuse of stale
access rights. %
With these hardware features in place, Okapi removes all universal
read gadgets~\cite{2019-McilroySevcik.etal} and provides security against breakout attacks
at thread level. %
\par
In addition, Okapi introduces a dedicated ISA instruction, called
\emph{OkapiReset}, which allows the programmer to revoke access rights explicitly, for example, after sensitive data regions have been
accessed legally. %
This allows a designer to prevent breakout attacks at a finer
granularity, e.\,g., at function level. %
\par
Okapi's protection mechanism guarantees that all data pages that cannot be read by any underlying software sandboxing mechanism will never be read speculatively by the microarchitecture either. This separation mechanism is orthogonal to existing in-process isolation techniques and efficiently closes their remaining gap for speculative execution.
\par
In our case studies~(\refsec{evaluation}), we demonstrate this for the state-of-the-art frameworks Wasmtime~\cite{2017-wasmtime} and ERIM~\cite{2019-Vahldiek-OberwangerElnikety.etal}.
\par
Besides breakout attacks, Okapi also mitigates poisoning
attacks~\cite{2023-NarayanGarfinkel.etal}. %
Okapi introduces efficient hardware features to drastically limit the
exploitable code space for Spectre gadgets. %
Whenever the instruction pointer speculatively crosses instruction
page boundaries, Okapi prevents speculative read accesses, which
inhibits the initial phase of a
Spectre-BTB~\cite{2019-KocherHorn.etal} attack. %
This measure prevents exploiting any gadgets outside the current trust
domain and restricts the set of potential gadgets for a poisoning
attack to only those that are located on the currently executing code
page. %
With this additional hardware mitigation in place, the software
measures elaborated in Sec.~\ref{sec:poisoning} offer a trade-off
between manual effort and performance to efficiently eliminate
poisoning attacks in a given application. %
\par
The rest of the paper is organized as follows: %
Sec.~\ref{sec:background} reviews background relevant for Okapi
followed by Sec.~\ref{sec:threat-model} which formalizes the security objective. %
In Sec.~\ref{sec:okapi} we introduce our Okapi concept and present its
security implications in Sec.~\ref{sec:security_reasoning}. %
Sec.~\ref{sec:architecture} describes the Okapi architecture in detail
before we evaluate different sandbox scenarios in a gem5 implementation
of Okapi in Sec.~\ref{sec:evaluation}. %
In Secs.~\ref{sec:HW-overhead} and~\ref{sec:security-evaluation}, we analyze an RTL implementation of the Okapi hardware w.\,r.\,t.\ its hardware overhead and security guarantees. %
In addition, Sec.~\ref{sec:security-evaluation} inspects the size of
trust domains in benchmark tests. %
Sec.~\ref{sec:sweet-spots} presents ``sweet spots'' for a practical application of Okapi and Sec.~\ref{sec:related-work} summarizes the state of the art of known
mitigations for TES attacks. %


\section{Background}
\label{sec:background}

\subsection{Spectre Variants}
\label{sec:spectre}
A TES attack~\cite{2019-CanellaVanbulck.etal} involves three main phases to transiently access sensitive data in memory that would not have been accessed in the sequential execution of the program: 

\begin{description}
\item[Setup] The adversary mistrains the targeted predictor and
  sets up the attacker process. %
  Additional preparation like flushing the caches helps to enlarge the
  time window for the transient execution phase. %
\item[Transient Access + Transmit] The victim code snippet is
  executed and transiently \textit{accesses} the secret memory
  region. %
  A subsequent transient operation \textit{transmits} the information
  by altering the microarchitectural state of the processor in a
  secret-dependent way. %
  After the core recognizes the misprediction, it rolls back to the
  last committed state, while some microarchitectural footprints
  (e.\,g., the cache state) remain in the system. %
\item[Secret Recovery] The attacker recovers the secret data
  from the microarchitectural footprint and makes it visible in the
  architectural state, often using timing-based cache side channels
  like Flush+Reload~\cite{2014-YaromFalkner} or a secret-dependent
  execution time in functional
  units~\cite{2020-FadihehMueller.etal}. %
\end{description}

Different types of Spectre attacks utilize different sources of speculative execution and therefore mainly vary in the second phase. 

Spectre-PHT~\cite{2019-KocherHorn.etal}, also known as Spectre variant~1, exploits the
pattern history table (PHT) that predicts the outcome of a branch. 
An example gadget is depicted in Fig.~\ref{fig:spectrev1}.
If the offset of the transient read instruction (\texttt{rbx} in Fig.~\ref{fig:spectrev1}) is controlled by the attacker, the instruction sequence serves as a so-called \emph{universal read gadget} that allows the attacker to read from the entire address space.

Spectre-BTB and Spectre-RSB exploit speculation on indirect branches (variant~2~\cite{2019-KocherHorn.etal}) and return addresses (variant~5~\cite{2018-MaisuradzeRossow, 2018-KoruyehKhasawneh.etal}). %
In this case, the predictors do not speculate on the outcome of the decision of a branch but on its target address. %
The attacker can poison the predictors to jump to an arbitrary
gadget even within the victim's code base to access the memory location of interest.
Pseudocode for a Spectre v2 attack is depicted in Fig.~\ref{fig:spectrev2}. %

Besides control flow predictions that are used in Spectre variants 1 and~2, data flow predictions can open transient execution side channels as well~\cite{2019-AMD,2018-Horn}.
Modern CPUs make predictions on target addresses within the load-store-unit to speed up data forwarding from one instruction to another. %
Not only does this pose a threat to data located in the store buffer, but it can also be used to construct a gadget that is able to transiently read from arbitrary memory addresses. %
An example of such a universal read gadget using data flow speculation is listed in Fig.~\ref{fig:spectrev4}. %
If the value of \texttt{x} is controlled by the attacker and the predictor is poisoned to predict the addresses \texttt{addr0} and \texttt{addr1} to match, the load-store-unit can forward the value of \texttt{x} to register \texttt{rax}. %
This value is then used as an address for the load instruction in line~3 to load from the attacker-controlled address \texttt{x} and afterward transmit the content of this memory address through a microarchitectural side channel. %

\begin{figure}[t]

   \begin{subfigure}{0.9\linewidth}
            \begin{lstlisting}[language=C, gobble=13]
              cmp rbx, ARR_LEN 
              jg out-of-bounds // mistrained pred.: not taken 
              mov r8, [rax + rbx] 
              mov r9, [r10 + 4096 * r8] // leak secret
              ...
             out-of-bounds:
              ...
        \end{lstlisting}
        \vspace{-4pt}
        \Description{Example pseudocode snippet of a Spectre-PHT gadget. An array access is guarded by an if-statement that ensures the array index to be in-bounds. If the outcome of the if-condition is predicted to be in-bounds, then the array access is performed and the returned value get's encoded into the cache to leak it.}
        \caption{Example pseudocode snippet of a Spectre-PHT gadget}
        \label{fig:spectrev1}
    \end{subfigure}
    \vspace{3pt}
    
    \begin{subfigure}{0.9\linewidth}
               \begin{lstlisting}[language=C, gobble=13]
              mov rbx, SECRET  // load secret into rbx
              jmp rax          // speculate rax=leak
              ...
             leak:             // encode secret in cache state
              mov rax, [B]
              mov rcx, [rax + rbx]
        \end{lstlisting}
        \vspace{-4pt}
        \Description{Example pseudocode snippet of a Spectre-BTB gadget. An attacker controlls the target prediction for indirecct jumps and can use this to speculatively divert the controll flow to a cache encoder gadget to leak the secret from a register.}
        \caption{Example pseudocode snippet of a Spectre-BTB gadget}
        \label{fig:spectrev2}
    \end{subfigure}
    \vspace{3pt}

    \begin{subfigure}{0.9\linewidth}
                   \begin{lstlisting}[language=C, gobble=13]
              mov [addr0], x       // attacker controlled x
              mov rax, [addr1]     // speculate addr0==addr1
              mov rdx, [rax]       // load secret from x
              mov rcx, [rbx + rdx] // leak secret
        \end{lstlisting}
        \vspace{-4pt}
        \Description{Example pseudocode of a Spectre-STL gadget. The attacker controlls an input value that is stored at some address addr0. If the CPU predicts addr0 to be equal to some addr1 that the code loads from, it speculatively forwards the attacker controlled value to the load instruction. If this loaded value is used as an index, the attacker can make the code speculatively read a secret value that can be leaked through, e.\,g., the cache.}
        \caption{Example pseudocode of a Spectre-STL gadget}
        \label{fig:spectrev4}
    \end{subfigure}
    \vspace{-4pt}
    \Description{Spectre code gadgets are are displayed in x86 pseudo-code. Subfigure a shows a Spectre-PHT gadget. Subfigure b shows a Spectre-BTB gadget. Subfigure c shows a Spectre-STL gadget.}
    \caption{Spectre code gadgets (x86 pseudo-code)}
    \label{fig:spectre}
\end{figure}

\subsection{Privilege Switches and Trust Domains}
\label{sec:context-switch}

\subsubsection*{Processes}
A \emph{process} is an instance of
a possibly multithreaded computer program that is being executed. %
Each process has an execution \emph{context}, consisting of an address space, a privilege level, and a list of threads. %
\emph{Threads} are independent sequential execution paths in a process that run concurrently if sufficiently many (logical) CPU cores are available.
The current hardware state of each thread consists of an instruction
pointer and the register contents. %
An \emph{address space} is a set of virtual addresses that are readable, writable, or executable by the threads of the process~\cite{2014-TanenbaumBos}.
The \emph{privilege level}
of a process defines its rights to modify other processes. %
 Processes with a higher privilege level are allowed to modify
processes with a lower privilege level but not vice versa. %
Such modifications include changes in the address space of a process
or even changes in register and memory values. %

\subsubsection*{Privilege Switch}
In the scope of this paper
we understand a \emph{privilege switch} as the change of the current
privilege level under which the currently active thread on a CPU
(logical) core executes.  %
Privilege switches happen as part of context switches, when switching between threads, as well as upon system calls from user-level
processes to privileged kernel code and their return back to user level. %
They are easily detectable in hardware. 

\subsubsection*{Sandbox}
A \textit{sandbox} refers to the (predefined) address space (instructions and data) that a program is confined to by a trusted handler, e.\,g., the operating system or a runtime environment. %

\subsubsection*{Trust Domains}
In this paper, a \textit{trust domain} denotes the currently executed code page together with the dynamically growing set of data pages that the currently running thread has, so far, accessed legally and that is therefore allowed to be accessed speculatively.
Depending on the chosen granularity of the sandboxes (e.\,g., threads, jobs, functions, etc.), 
trust domains can grow to different sizes.
When handing over control from one sandbox to another all access permissions are reset.
This ensures that sandboxes are prevented from accessing each other's data pages. %

\subsection{Virtual Memory and Paging}
\label{sec:VM}
 Most modern computer systems employ virtual memory (VM)~\cite{2017-BhattacharjeeLustig}, providing the user with an
abstract programming interface. %
The key concept is to use virtual addresses that are translated by the
VM system to physical addresses to access the requested data. %
Virtual memory is organized in \emph{pages}\footnote{One page is
  typically 4\,KB in size. %
  Modern VM systems usually also support 2\,MB or~1\,GB pages which
  are often called \emph{hugepages} or \emph{megapages}.} that are
mapped to physical \emph{page frames} of the same size. %
The mapping for each process is stored in a separate \emph{page table}
that is managed by the operating system. %
If a process accesses a virtual address, the VM system performs a
\emph{page table walk} to translate the virtual address to a physical
address. %
Page table walks are considered to be slow as they involve several
accesses to physical memory to look up the mapping. %
Additionally, VM annotates pages with permissions (read, write,
execute), thereby implementing simple memory protection and
isolation. %
The VM subsystem is crucial for system performance since about one
third of fetched instructions interact with the memory
system~\cite{2017-BhattacharjeeLustig}. %
Our own experiments with the SPEC CPU2017 benchmark suite, in which up
to~25\,\% of instructions are \emph{loads}, support this statement (cf.~\refsec{SPEC}). %
Therefore, computer architects introduce dedicated hardware to speed
up address translation. %
The Translation Look-aside Buffer (TLB) is a low-latency cache that
keeps track of recent address translations to avoid costly page table
walks. %

\section{Security Objective}
\label{sec:threat-model}

The security objective of Okapi is to prevent leakage of 
data residing in data memory or in the cache hierarchy, i.\,e.,
\emph{data at rest}. %
Similar to existing hardware mitigations against transient execution
side channels, like STT~\cite{2019-YuYan.etal}, Eager/Na\"ive
Delay~\cite{2019-SakalisKaxiras.etal}, and
NDA~\cite{2019-WeisseNeal.etal}, Okapi targets the objective of
\textit{weak speculative
  non-interference}~\cite{2021-GuarnieriKoepf.etal}. %
\par

Following the terminology of~\cite{2021-GuarnieriKoepf.etal}, we denote a memory location as \emph{high} (H) if it stores confidential data, or as \emph{low} (L) if the data at the location is public. %

\paragraph{Breakout Security}
We assume that a program~$p$ is sandboxed in software (\emph{vanilla sandboxed}), i.\,e., the high memory locations do not influence the architectural state under ~$p$'s sequential program semantics. %
The program~$p$ is breakout-secure or \emph{generally sandboxed} if it also satisfies \emph{weak speculative non-interference}. %
This means that the high locations do not influence the architectural state even under speculative semantics~\cite{2021-GuarnieriKoepf.etal}. %
Provided that high and low locations do not share memory pages (i.\,e., software sandboxes are implemented page-aligned), Okapi is guaranteed to enforce breakout security for~$p$, as described in Sec.~\ref{sec:breakout}. 

\paragraph{Poisoning Security}
In contrast to breakout attacks, poisoning attacks target a victim
program~$p$ that has access to high memory locations. %
Additionally, $p$ has measures in place to protect high memory
locations from being leaked sequentially, e.g., by relying on constant-time code for sensitive computations. %
Nonetheless, an attacker can poison the control flow in~$p$ to
circumvent these measures under speculative semantics and leak
confidential data~\cite{2022-CauligiGuarnieri.etal,
  2021-NarayanDisselkoen.etal}. %
\par %
The victim program can be partitioned into two sets of sections, $p_H$ and~$p_L$. %
All sections of~$p$ accessing high memory locations sequentially
belong to~$p_H$, while~$p_L$ is the complement of~$p_H$. %
In other words, ${p_L}$ 
never accesses confidential data and
is 
therefore
\emph{vanilla sandboxed}~\cite{2021-GuarnieriKoepf.etal} w.\,r.\,t.\ the high memory locations of~$p$. %
The program must be partitioned in a way such that each
individual section of~$p_H$ cannot leak more than the data it
accesses sequentially. %
Under such a partitioning, $p$ is poisoning-secure iff ${p_L}$ is
breakout-secure w.\,r.\,t.\ the high memory locations. %
A trivial partitioning selects~$p_H$ to contain only all load
instructions to high locations. %
Sec.~\ref{sec:poisoning} addresses the measures Okapi provides for
poisoning security and discusses how to find an efficient and
effective partitioning of~$p$. %
\par
We consider all types of speculation that can open a transient window for a load instruction, including control-flow predictions, exceptions, memory-dependence speculation, and memory consistency violations. %
\par
Guarantees for \textit{speculative non-interference}~\cite{2021-GuarnieriKoepf.etal}, i.\,e., ensuring that transient execution side channels do not reveal data in the register file and instruction operands (data in transit), are outside the scope of this paper. %
This includes side channels due to predictive store forwarding~\cite{2021-AMD} and other optimizations that can leak the content of the store buffer. In addition, we exclude classical side channel attacks, e.\,g., monitoring the instruction cache footprint of square-and-multiply exponentiation in RSA~\cite{2014-YaromFalkner} or data-dependent prefetcher-based attacks violating constant-time guarantees~\cite{2024-ChenWang.etal}. %

Moreover, physical side channels, including power or electromagnetic side channels, fall outside the scope of this paper. %
\par
Our security objective covers all TES attacks initialized by speculative loads, independent of the microarchitectural side channel being used to transmit and recover secrets (e.\,g., cache side channels~\cite{2019-KocherHorn.etal}, port contention~\cite{2021-BehniaSahu.etal}, or data-dependent timing in functional units~\cite{2015-AndryscoKohlbrenner.etal}). %
In particular, it covers the formidable Spectre-PHT and Spectre-BTB attacks for which, to date, no satisfactory mitigations exist in industrial practice. %


\section{Okapi Concept}
\label{sec:okapi}
Common to all Spectre variants targeting data at rest is a speculative load instruction that accesses a memory location that could not be accessed by the sequential program flow. %
Preventing the execution of this transient access instruction is sufficient to avert an attack. %
Blocking all speculative loads from executing would achieve this goal but comes with a significant performance penalty, as shown in~\cite{2023-JauchWezel.etal}. %
Most of the load instructions executed speculatively are not part of an attack. %
For minimum performance penalty, it is therefore crucial to only block the malicious ones.

Okapi builds upon this observation by blocking as few speculative load
instructions as possible, taking into account legal accesses by the current trust
domain.
Our proposed architecture leverages the spatial and temporal locality
in memory accesses, which means that most of the data retrieved within
a short period of time is located in a narrow region in the memory space. %
Fig.~\ref{fig:spec-cx-switch-pages} shows, for the SPEC CPU2017 rate
suite, that programs indeed fulfill this locality assumption as they access
only about 20 pages between context switches, on average. %
In fact, all benchmarks access at most 60 pages per context switch
except for the first context switch, during which around 300 pages are accessed to load
the binaries into memory. %

In contrast to other isolation mechanisms, such as \emph{in-process
  isolation}~\cite{2019-Vahldiek-OberwangerElnikety.etal,2020-SchrammelWeiser.etal},
Okapi is not aware of a program's address space and sandbox boundaries.
Instead, it supplies a secure hardware basis that guarantees the enforcement of boundaries provided by security mechanisms in software.
The Okapi hardware learns about legal accesses by tracking committed load instructions during the execution. %
Thus, an Okapi trust domain comprises the (dynamically growing) set of pages that the currently running thread has, so far, accessed legally and that it may, therefore, also access speculatively. %

\begin{figure*}[t]
    \centering
    \includegraphics[width=0.9\linewidth]{fig/spec2017_rate_cx-switch_data.png}
    \vspace{-1em}
    \Description{
        The figure shows a bar plot. On the x-axis, most of the SPEC CPU2017 benchmark programs are listed. The y-axis displays the number of accessed memory pages these programmes accessed. Per program, four bars are given: the total number of code pages, the total number of data pages, the code pages mean per context switch, and the data pages mean per context switch. On average, each programm access a total number of around 280 code pages and around 24 data pages. Per context switch, however, the each program only accesses around 18 code pages and around 8 data pages.}
    \caption{
        The average number of accessed pages (code and data) in between context switches and the total number of allocated pages for the SPEC CPU2017 rate suite.}

    \label{fig:spec-cx-switch-pages}
\end{figure*}

\subsection{Achieving Breakout Security}
\label{sec:breakout}
Okapi allows speculative access only to memory pages that have already been accessed non-speculatively at least once by the same sandbox. %
This confines potential transient attacks to only the pages belonging to the current trust domain and removes all universal read gadgets capable of speculatively leaking data at arbitrary memory addresses by breaking out of the trust domain.

Okapi enforces speculative read permissions with a single \textit{safe access bit} added to each page table entry in the data TLB. %
Whenever a page is accessed non-speculatively, the bit is set. %
Speculative loads are only allowed to obtain their physical address from a TLB entry if the corresponding \textit{safe access bit} is set. %
Otherwise, the address translation is blocked until the load becomes non-speculative. %
This measure effectively prevents Spectre gadgets from accessing memory locations that do not have the \textit{safe access bit} set and thwarts breakout attacks for page-aligned sandboxes. %
In many cases, the first access to a page causes a miss in the TLB and triggers an expensive page table walk. %
Hence, the delay induced by this mechanism in Okapi is largely shadowed by the page table walk which reduces the overall performance overhead incurred by Okapi.

In this paper, we describe the Okapi mechanism for a single core without multi-threading, featuring a physically indexed, physically tagged cache. For other system configurations, including ones with simultaneous multi-threading (SMT), the functionality can be adapted accordingly. For example, a separate \textit{safe access bit} can be implemented per SMT thread.

Note that data-dependent prefetchers need to perform an address translation for dereferencing potential pointers~\cite{2022-VicarteFlanders.etal}, i.e., the prefetcher requests can be blocked with the same mechanism in the TLB.
Thus, data-dependent prefetchers do not pose an additional breakout threat to data at rest~(cf.~\refsec{threat-model}).

\paragraph{Resetting the Permissions}
During execution, the program successively accesses data pages architecturally and sets the corresponding \emph{safe access bits}. %
Accordingly, the trust domain, i.\,e., the number of memory pages from which data can possibly be leaked transiently, grows over time. %
Resetting the \textit{safe access bits} at the right point in time is crucial to prevent the (mis-)use of stale access rights. %
The size of a sandbox that establishes a trust domain can vary significantly, ranging from single functions to entire threads or processes. 
For correct handling of the access bits in Okapi, transitions between trust domains must be transparent to the CPU. %

Sandboxes at process or thread level can easily be implemented in hardware. %
Privilege switches (cf.~Sec.~\ref{sec:context-switch}) are detected by monitoring the corresponding Control and Status Registers (CSRs) and automatically clear the \textit{safe access bits}. %

Therefore, as a first important result, we note that the Okapi hardware alone inhibits all universal read gadgets (e.\,g., the ones exploited by Spectre-PHT, -BTB and -STL, as depicted in \reffig{spectre}) and achieves breakout security for thread-sized sandboxes. %

In theory, to prevent breakout attacks one could create a new process for every sandbox.
In practice however, it is desirable to have more than one sandbox within the same thread or process to avoid the high cost of context switches~\cite{2024-cloudflare-workers}. %

When multiple sandboxes are mapped to the same process, the straightforward approach of delaying load instructions in case of a TLB miss~\cite{2019-SakalisKaxiras.etal} is not sufficient to guarantee secure speculation.
The first sandbox populates the TLB with its entries and subsequent sandboxes can re-use these entries to perform a breakout attack. 

While sandboxes are clearly distinguishable in the software, detecting transitions between sandboxes below thread level is not directly possible in the hardware. %
For this reason, Okapi introduces a special instruction using which the software can communicate sandbox exits to the hardware. %

\paragraph{OkapiReset Instruction}
We add a dedicated instruction, \emph{OkapiReset}, to the ISA that resets the trust domain by flushing all \textit{safe access bits}. %
The \emph{OkapiReset} instruction acts as a speculation barrier for load instructions. %
Similar to other speculation barriers, such as fences, it is executed only upon reaching the head of the Re-Order Buffer (ROB) to avoid premature resetting. %
As long as an \emph{OkapiReset} instruction is present in the ROB, all younger load instructions are serialized to prevent out-of-order use of existing access permissions. %
Software developers can make use of the \emph{OkapiReset} instruction to implement more fine-grained sandboxes, e.\,g., at function level. %

In conclusion, the combination of the Okapi hardware and the \emph{OkapiReset} instruction enforces breakout security for custom-sized sandboxes below thread level. %

\subsection{Achieving Poisoning Security}
\label{sec:poisoning}
While establishing trust domains for programs of a chosen granularity prevents transient accesses to the outside, data located on pages with an activated \textit{safe access bit} can still be loaded speculatively. %
The challenge for Okapi w.\,r.\,t.\ poisoning attacks (cf.~Sec.~\ref{sec:threat-model}) is to guarantee that code sections~$p_L$ within a program that do not sequentially access confidential data cannot misuse speculative access permissions gained by sensitive code sections~$p_H$. %
We need to ensure that upon a transition from~$p_H$ to~$p_L$ the high memory locations are not contained in the trust domain. %
Placing confidential and public data on separate pages and augmenting the program with an \emph{OkapiReset} instruction after every access to confidential data solves this problem as it effectively constrains $p_H$ to contain all confidential load instructions. 
However, this incurs a substantial performance overhead as every \emph{OkapiReset} instruction revokes all \emph{safe access bits}, thereby also preventing speculation on previously accessed public data. %
\par%
We therefore introduce the \emph{OkapiLoad} instruction that allows the software to access confidential data pages non-speculatively without adding the page to the trust domain, i.\,e., not setting the \emph{safe access bit}. %
This ensures poisoning security (cf.~Sec.~\ref{sec:threat-model}) if every access to confidential data is performed by an \emph{OkapiLoad} and preserves speculative access rights on previously accessed public data.

Achieving full poisoning security for sandboxes that repeatedly process confidential data with these instructions comes at a high overhead due to serializing accesses to high memory locations. %
This can be relaxed by increasing the size of the sections in~$p_H$ from single load instructions to larger segments of code to be able to speculate on high memory locations without leaking them.

A major challenge for increasing the size of the sections in~$p_H$ are Spectre-BTB attacks, illustrated in Fig.~\ref{fig:okapi_v2}. %
By poisoning branch prediction, an attacker can transiently divert the control flow from the sequential program flow (green arrow) to a gadget (blue arrow). %
Without additional measures, transiently executed gadgets can still access (red arrow) the content of pages already marked with a \textit{safe access bit}, i.\,e., data pages accessed by~$p_H$. %
This gadget can reside anywhere in the mapped code space, posing a challenge to conventional mitigations~\cite{2024-WiebingTron.etal}.

Therefore, we introduce an additional low-cost countermeasure directly in the hardware that drastically reduces the remaining gadget space. %
Okapi achieves this by delaying the execution of loads following a speculative control-flow decision crossing instruction page boundaries. %
Such loads can be detected in the fetch stage of the processor. %
If the next instruction to be fetched speculatively is located on a different page than the current instruction, the next instruction is flagged as \textit{suspicious} (flag symbol in Fig.~\ref{fig:okapi_v2}). %
During execution, every load instruction following a \textit{suspicious} instruction is delayed until the control flow of the \textit{suspicious} instruction becomes non-speculative. %
In Fig.~\ref{fig:okapi_v2} this is illustrated by Okapi intercepting the malicious read access. %
This measure efficiently prevents poisoning attacks that speculatively cross an instruction page boundary.

With this hardware measure in place, poisoning attacks are only possible if the control flow can (speculatively) reach a gadget without crossing page boundaries and the gadget targets confidential data on pages that are already a part of the trust domain, i.\,e., that have their \textit{safe access bit} set. %
This facilitates the task for the software developer to protect against poisoning attacks at larger granularity (and, hence, lower performance cost).

We can enlarge the sections in~$p_H$ as long as they fulfill the property of not containing any gadgets (cf.~Sec.~\ref{sec:threat-model}). %
Therefore, if we can guarantee a candidate region to be free of gadgets~\cite{2024-WiebingTron.etal}, we can extend the sections in~$p_H$ to a chosen granularity, e.\,g., an entire function. %
The inspection of candidate regions can be done independently of each other. %

In conclusion, Okapi offers two different strategies to blocking poisoning attacks in sandboxes. %
The developer can either prevent confidential data pages from entering the trust domain by using the \emph{OkapiLoad} instruction or revoke the access rights via an \emph{OkapiReset} instruction when switching from~$p_H$ to~$p_L$. %

Note that the hardware measures alone drastically reduce the code space available for exploitable gadgets as shown in our experiments in Sec.~\ref{sec:attack-surface}.

Applications that rarely access confidential data might benefit from guarding these accesses with \emph{OkapiLoad} instructions and keeping the speculative access rights on public data for~$p_L$. %
On the other hand, applications that repeatedly access confidential data benefit from larger code sections in~$p_H$. %

\subsection{Okapi's Security Guarantees}
In summary, the features described in \refsec{breakout} and \refsec{poisoning} enable Okapi to effectively and efficiently protect against breakout and poisoning attacks.
In particular, Okapi eliminates all threats imposed by Spectre-like universal read gadgets with only minor performance and hardware penalty (cf.~\refsec{evaluation} \& \refsec{HW-overhead}).  
Note that this security guarantee is independent of the classification of different Spectre variants.
Okapi prevents unsafe access instructions which are a necessary component of every Spectre attack, independent of the covert channel being used.
Hence, it covers all Spectre variants targeting data residing in memory.
\par
In addition, Okapi by design eliminates Meltdown- and MDS-style vulnerabilities since it permits access to physical memory addresses only in case of a non-faulting virtual address translation~(cf.~\refsec{security-evaluation}).
Thus, attackers cannot bypass any kind of memory exception which is the underlying mechanism for these attacks.
\par
The hardware features incur low performance and implementation overhead and provide breakout security for thread-sized sandboxes (cf.~\refsec{formal}). %
In addition, without further measures, they already drastically limit the set of possible locations for gadgets exploitable in poisoning attacks. %

With additional security measures at the software level leveraging the \emph{OkapiReset} instruction, Okapi delivers breakout security in custom-sized sandboxes (cf.~\refsec{mutually-distrusting}). %
Full security against poisoning attacks can be achieved by identifying and hardening gadgets solely within the limited code region defined by the victim sandbox's code. %
Alternatively, access to secret data can be controlled using the \emph{OkapiLoad} instruction. This, at the cost of additional performance overhead, relieves the software developer from manually identifying gadgets within the victim's code base (cf.~\refsec{vaulting}). 
 \par
Both Okapi instructions act like pseudo-fences.
They serialize younger load instructions and thus do not cause any additional data-dependent timing. 
Furthermore, load instructions in Okapi only access addresses on non-faulting pages and do therefore not create any secret-dependent behavior.
As a result, Okapi does not compromise any existing constant-time guarantees.

\begin{figure}[t]
    \centering
    \includegraphics[width=.7\linewidth]{fig/OkapiV2Figure.png}
    \vspace{-1em}
    \Description{
        This figure shows three data pages that currently belong to the trust domain and contain secret data. Also, two code pages are show: one contains the code that is currently being executed and contains an indirect jump, and one page that contains a Spectre gadget. Arrows indicate Okapis behavior when an attacker poisons the branch predictor to divert the indirect jump to speculatively execute the Spectre gadget on the second code page. Since the gadget lies on a different page, every load performed by the gadget gets flagged as suspicious and gets withheld until its execution becomes save. This does not happen since the gadget does not execute architecturally and therefore Okapi effectively blocks the gadget's access to the secret.
    }
    \caption{
        Spectre-BTB challenge for Okapi. If speculative execution flow leaves the current page, all loads on the \emph{suspicious} page are delayed until accesses to that code page are no longer speculative.
    }
    \vspace{-1em}
    \label{fig:okapi_v2}
\end{figure}


\section{Okapi in different threat scenarios}
  
\label{sec:security_reasoning}
Okapi prevents speculative leakage of data outside a program's memory space. %
Compared to schemes that enforce \emph{weak speculative non-interference} at an address granularity~\cite{2019-YuYan.etal,2019-SakalisKaxiras.etal,2019-WeisseNeal.etal}, Okapi achieves significantly better performance by declassifying an entire data page with the first non-speculative access to this page. %
As long as no \emph{OkapiLoad} instruction is used, the Okapi hardware dynamically labels sequentially accessed data pages as low, i.\,e., public. %
This allows for a highly efficient protection against cross-sandbox breakout attacks solely in hardware. %
\par
As a result of the native hardware support for virtual memory translation based on page tables (cf.~Sec.~\ref{sec:VM}), most operating systems and runtime environments, such as in-process isolation~\cite{2020-SchrammelWeiser.etal,2023-NarayanGarfinkel.etal} and enclaves~\cite{2019-BourgeatLebedev.etal}, utilize page granularity to define the accessible address space and enforce software boundaries. %
Commercial cloud services, such as Cloudflare workers~\cite{2024-cloudflare-security}, similarly incorporate page-aligned partitioning of the address space between mutually distrusting tenants. %
Therefore, mutually distrusting software domains typically possess non-overlapping page-aligned data memory address spaces, and page sharing only happens in instruction memory through shared libraries. %
Thus, in the context of sandboxing, secure speculation with finer than page granularity has limited benefits while incurring a substantial performance penalty. %
\par
In the following, we discuss different scenarios for environments that benefit from the security guarantees provided by Okapi: %

\subsubsection*{``Classic'' Sandboxing}
In classic sandbox scenarios where a trusted environment is protected from being accessed by an untrusted instance, we only need to ensure that all \textit{safe access bits} are reset when handing over control to the untrusted job. %
If switching to the untrusted job involves a privilege switch, this is already taken care of by the Okapi hardware. %
Otherwise, the sandbox runtime must execute an \textit{OkapiReset} instruction to revoke all previously gained access rights. %

\paragraph*{Mutually Distrusting Sandboxes}
Securing runtime environments that run multiple tenants within a single thread against breakout attacks is a major concern in industrial applications, e.\,g., serverless computing~\cite{2022-SchwarzlBorello.etal, 2024-cloudflare-security}. %
Okapi secures this kind of environment by preventing speculative cross-domain accesses even within the same virtual address space. %
Therefore, the runtime environment, e.\,g., the WebAssembly engine, needs only minor modifications to execute an \textit{OkapiReset} instruction each time it enters and exits a sandbox, to fully separate the runtime and each sandbox from each other. %

\subsubsection*{Vaulting}
The inverse scenario to ``classic'' sandboxing is vaulting, where a program is protected against malicious accesses from the outside. %
Besides breaking into the vault, attackers can launch a poisoning attack (cf.~Sec.~\ref{sec:threat-model}) to trick the victim program into speculatively leaking its own confidential data. %
Okapi already covers breakouts from other programs into vaulted code segments with the security mechanisms protecting mutually distrusting sandbox environments. %
In addition, Okapi mitigates poisoning attacks by employing the hardware measures described in Sec.~\ref{sec:poisoning}. %
If desired, the \textit{OkapiLoad} instruction prevents confidential data from being added to the vault's trust domain. %

\subsubsection*{Kernel as a Confused Deputy}
The Okapi hardware automatically resets the \emph{safe access bits} on every privilege switch independent from the software. Therefore, confused-deputy attacks against the kernel are automatically limited. 
As soon as the privilege switch occurs, the kernel can no longer speculatively access data pages belonging to the threads that were running before.
Therefore, the kernel can never leak a thread's information that it does not access architecturally during a privilege switch.
\par
If the attack targets data that is accessed by the kernel sequentially, it can only be successful if there is a gadget on the same instruction page of the kernel code.


\section{Okapi Architecture}
\label{sec:architecture}
\subsection{Hardware Architecture}
\label{sec:HWarchitecture}
\begin{figure}[t]
    \centering
    \includegraphics[width=0.9\linewidth]{fig/Okapi2.png}
    \vspace{-1em}
    \Description{
        The figure shows a 6 stage CPU pipeline. The stages are: fetch, decode, rename, execute, writeback, and commit. The fetch stage has access to the instruction cache port. After the rename stage, the instruction is forwarded to the execute stage, the re-order buffer (ROB), the instruction queue (IQ), and the load store queue (LSQ), which in turn has access to the translation-lookaside buffer (TLB) with the safe access bits and the data cache port. Some components are marked with numbers that are explained in the text. The fetch stage is marked with one, decode with two, the ROB with three, the LSQ with four, the TLB with five, the IQ with six and the execute stage with seven.
    }
    \caption{
        Out-of-Order processor augmented with Okapi.
    }
    \label{fig:Okapi_gem5}
\end{figure}

Okapi can be implemented in an out-of-order CPU as shown in Fig.~\ref{fig:Okapi_gem5}. %
The modified parts are marked with numbers. %

\begin{enumerate}
\item
In the fetch stage every \textit{\{PC, next PC\}} pair is compared to detect instructions that cross page boundaries. %
If the next PC is located on a different page the instruction is marked as \textit{suspicious} for a possible Spectre-BTB/RSB attack. %

\item
The decode stage is extended to handle the \emph{OkapiReset} and \emph{OkapiLoad} instructions. %

\item
The ROB is modified to keep track of the visibility point, i.\,e., the youngest instruction in terms of program order that can open a transient window (cf.~Sec.~\ref{sec:threat-model}). %
Load instructions inside the ROB are marked as \textit{unsafe} upon their insertion if they are younger than the visibility point and thus not certain to commit. %
An \emph{unsafe} load instruction becomes \emph{safe}, when it is passed by the visibility point, i.\,e., it is bound to commit. %
 Similarly, loads are marked as \textit{suspicious\_load} if they are younger than an unsafe \textit{suspicious} instruction. %
Loads between the visibility point and the first \textit{suspicious} instruction clear their \textit{suspicious\_load} bit. %

\item
The Load Store Queue (LSQ) selectively blocks loads that are marked with the \textit{suspicious\_load} bit from going to the TLB and thus from executing speculatively. %

\item
Inside the TLB, \textit{safe} loads behave as usual and set the \textit{safe access bit} for the corresponding page upon a successful translation, while speculative and therefore \textit{unsafe} loads can only be translated if the \textit{safe access bit} for the corresponding TLB entry is already set. %
\textit{OkapiLoads} behave analogously except for never setting the \emph{safe access bit}. %
\textit{Privilege switches} are detected by monitoring the CSRs storing the current privilege level. %
The \textit{safe access bits} are cleared each time the privilege level changes. %
In addition, the TLB is modified to reset the \textit{safe access bits} when the \emph{OkapiReset} instruction is executed. %
\item
The Instruction Queue (IQ) of the processor is extended to reschedule all blocked load instructions. %
This includes the blocked \textit{suspicious\_load} instructions from the LSQ as well as \textit{unsafe} loads that did not succeed in the TLB. %
All loads are rescheduled as soon as they are marked \textit{safe} within the ROB while \textit{suspicious\_load} instructions can be rescheduled speculatively as soon as the \textit{suspicious\_load} bit is cleared. %

\item
The \emph{OkapiReset} instruction is delayed until it reaches the ROB head, ensuring that all previous loads have been executed. %
The loads following the \emph{OkapiReset} instruction are delayed until they become non-speculative or until the \emph{OkapiReset} instruction is executed. %
This prevents the re-use of stale permissions. %
\end{enumerate}

The majority of the described extensions require only minor changes to the pipeline, such as adding single bits to micro-operations or TLB entries. %
Okapi, to a large extent, re-uses the existing logic in the LSQ and requires no modification to the cache and memory subsystem. %
Implementing the Okapi instructions and the logic to determine the visibility point are the only significant hardware extensions. %

\subsection{Additional Software Support}
\label{sec:software}
As described in Sec.~\ref{sec:okapi}, the Okapi hardware depends on additional software measures to effectively eliminate breakout attacks below thread level, and poisoning attacks. %
As a summary, the following measures taken at the software level 
are required to ensure the security guarantees: %
 \begin{itemize}
    \item Confidential data must not be placed on the same page as public, regularly accessed data. %
    Thereby, only accesses to the confidential data pages need to be safeguarded and unnecessary slowdown in the non-sensitive part of the code is avoided. %
    \item Transitions from one software sandbox below the size of a thread to another must perform an \emph{OkapiReset}. %
    \item Transitions from critical code sections in~$p_H$ to non-sensitive code sections in~$p_L$ must perform an \emph{OkapiReset}. %
This requirement can be relaxed if confidential data is only being accessed using \emph{OkapiLoad}. %
    \item The sizes of the critical code sections in~$p_H$ can be increased to, e.\,g., functions, if the instruction pages of these sections are known to not contain any gadgets. %
  \end{itemize}

The trusted entity, e.\,g., the operating system, needs to provide page alignment for the data of each sandbox. %
The software developer can implement the above measures by annotating secret data and using an extended compiler to insert \emph{OkapiReset} and/or \emph{OkapiLoad} instructions accordingly. %
In case the user wishes to increase the sections in~$p_H$, the absence of gadgets can be verified by manual inspection or gadget analysis tools, e.\,g.~\cite{2024-WiebingTron.etal, 2020-googlesecurity}. %


\section{Performance Evaluation}
\label{sec:evaluation}
The proposed architecture is implemented\footnote{Available on GitHub: \url{https://github.com/RPTU-EIS/gem5-Okapi}} and evaluated based on the
out-of-order CPU model \emph{o3 CPU} provided by the gem5
simulator~\cite{2011-BinkertBeckmann.etal}, configured according to
Tab.~\ref{tab:gem5_config}. %
First, we evaluate the performance of Okapi after implementing only
the hardware features. %
This determines the overhead for enforcing breakout security in thread-sized sandboxes and limiting the gadget space for poisoning attacks to code pages that are accessed by the sequential data flow (Sec.~\ref{sec:SPEC}). %
Next, we showcase how different jobs within the same thread of a WebAssembly runtime can easily be hardened
against sandbox breakout attacks using the \emph{OkapiReset} instruction (Sec.~\ref{sec:wasmtime}). %
Afterwards, we showcase how the MPK-based sandboxing scheme ERIM~\cite{2019-Vahldiek-OberwangerElnikety.etal} can be adapted to Okapi (Sec.~\ref{sec:erim}).
Lastly, we examine the performance overhead caused by using the \emph{OkapiReset} and \emph{OkapiLoad} instructions to explicitly vault sensitive data in a cryptographic library (Sec.~\ref{sec:vaulting}). %

\begin{table}[t]
    \footnotesize
    \centering
    \caption{gem5 configuration}
    \label{tab:gem5_config}
    \vspace{-1em}
    \resizebox{0.8\linewidth}{!}{%
    \begin{tabular}{lr}
        \toprule
        Processor                   &                           \\
        \midrule
        ISA                         & x86\_64                   \\
        Cores              & 1                         \\
        Decode width                & 5                         \\
        Issue/Commit width          & 8                         \\
        ROB entries                 & 192                       \\
        Instruction queue entries   & 64                        \\
        LD/ST queue entries         & 32/32                     \\
        \bottomrule
        \toprule
        Memory                      &                           \\
        \midrule
        DTLB entries                & 64                        \\
        L1 I cache                  & 32\,KiB, 8 ways           \\
        L1 I access latency         & 6 cycles roundtrip        \\
        L1 D cache                  & 48\,KiB, 12 ways          \\
        L1 D access latency         & 6 cycles roundtrip        \\
        Private L2 cache            & 1280\,KiB, 20 ways        \\
        Private L2 access latency   & 60 cycles roundtrip       \\
        Main memory                 & 16\,GB DDR4 2400\,MHz 8x8 \\
        \bottomrule
    \end{tabular}
    }
    \vspace{-0.5em}
\end{table} 

\subsection{Thread-Level Sandboxing}
\label{sec:SPEC}

\begin{figure}[t]
    \centering
    \includegraphics[width=0.95\linewidth]{fig/okapiSpec2017_one_col.png}
    \vspace{-1em}
    \Description{
        This figure shows the SPEC CPU2017 performance results of Okapi and three countermeasures that are discussed in the literature. These countermeasures are Na\"ive Delay and Eager Delay described in~\cite{2019-SakalisKaxiras.etal}, and dynamic information flow tracking (DIFT) similar to~\cite{2019-YuYan.etal, 2021-LoughlinNeal.etal}. The performance results are given as normalized execution time over the insecure baseline design. The results show that Okapi incurs about 3.17\,\% overhead in the geometric mean and outperforms all other approaches for nearly all benchmark programms. DIFT is the runner-up with 18.36\,\% overhead on average.
    }
    \caption{
        SPEC CPU2017 performance results
    }
    \label{fig:Okapi_benchmarks2017}
    \vspace{-1em}
\end{figure}

We run the SPEC~CPU2017~\cite{2018-BucekLange.etal} reference workloads in full system mode using SimPoints~\cite{2005-HammerlyPerelman.etal} to benchmark the gem5 Okapi design. %
The first 100 billion instructions are profiled to create a maximum of five SimPoints comprising 100 million measurement and~50 million warm-up instructions for each workload. %
The evaluation considers only the workloads that we were able to run on the unmodified o3 CPU model. %
\par
For comparison, in addition to Okapi, we also implemented other hardware-based architectures that mitigate TES attacks targeting data at rest: %
\begin{itemize}
    \item Na\"{i}ve delay~\cite{2019-SakalisKaxiras.etal} -- executes loads once they reach the head of the ROB. %
    \item Eager delay~\cite{2019-SakalisKaxiras.etal} -- executes load instructions as soon as they are no longer speculative. %
    \item A dynamic information flow tracking (DIFT) architecture similar to~\cite{2019-YuYan.etal, 2021-LoughlinNeal.etal}\footnote{
    STT~\cite{2019-YuYan.etal} reports numbers based on SPEC~CPU2006. %
The provided code does not run on the latest versions of gem5. %
Therefore, we ported STT to gem5 23.0.0.1, so that all the compared schemes run the same checkpoints of the most recent SPEC~CPU benchmarks on the same simulator. %
    The performance overhead reported by the authors of STT using the same sources of speculation is~14.5\,\%.} -- 
In DIFT, speculative load instructions are allowed to access the memory and initiate a taint for the accessed data. %
Speculative instructions that are able to form side channels are blocked from using tainted data. %
\end{itemize}

For the experiments reported in this subsection neither \emph{OkapiReset} nor \emph{OkapiLoad} instructions were inserted into the benchmarks. %
The performance results for the SPEC~CPU2017 suite are depicted in Fig.~\ref{fig:Okapi_benchmarks2017}, summarizing the runtime overhead normalized to the insecure baseline design. %
It can be seen that Okapi outperforms all other mitigation schemes with an average performance overhead of only 3.17\,\%. %
The easy-to-implement blanket fixes, na\"{i}ve and eager delay, incur a delay of  
162.03\,\% or~98.44\,\%
, respectively. %
While DIFT requires more complex modifications to the processor's hardware (cf.~Sec.~\ref{sec:related-work}), it still imposes a performance overhead of 18.36\,\%. %

The significant performance advantage of Okapi originates in its nature of adding an entire data page to the trust domain at a time. %
This means that it does not need to delay loads to the same page twice. %
The other examined fixes enforce \emph{weak speculative non-interference} at address level. %
However, all architectures provide the same level of breakout security if the thread's data is page-aligned. %

\par
In the following, we analyze the results for the Okapi architecture in more detail. %
The vast majority of the load instructions enters the ROB as potentially malicious, as depicted in Fig.~\ref{fig:ROBLoads} Appendix~\ref{sec:appendix_experiments}. %
On average, 85.5\,\% of the dispatched load instructions commit their results to the architectural state. %
However, this varies from 60.7\,\% in the case of \emph{mcf} to~99.9\,\% for \emph{cactuBSSN}. %
Fig.~\ref{fig:OkapiLoads} in Appendix~\ref{sec:appendix_experiments} illustrates the status of speculative load instructions when the LSQ first tries to execute their address translation. %
A load instruction can either be blocked due to being marked as \emph{suspicious\_load} or be passed to the TLB. %
On average, around 14.3\,\% of the speculative load instructions are blocked from going to the TLB due to our countermeasure to prevent the use of Spectre-BTB gadgets. %
Out of the remaining load instructions 98.4\,\% have their \emph{safe access bit} set, confirming our locality assumption. %
\par
In general, workloads perform well if they have a high locality in combination with only few load instructions being blocked due to being marked as \textit{suspicious\_load}. %
This can be observed at the examples of \emph{bwaves} and \emph{lbm}. %
Delayed \textit{suspicious\_load} instructions do not always have a high impact on performance because they can be rescheduled speculatively as soon as the control flow speculation resolves. %
This can be observed at the example of \emph{perlbench\_1} where 42.1\,\% of the speculative load instructions are not allowed to go to the TLB, initially. %
However, this workload still performs at the same speed as the insecure baseline. %
\par
On the other hand, in \emph{cactuBSSN}, for example, 99.8\,\% of the dispatched load instructions commit, while only 91.3\,\% of the issued and unsafe load instructions have the \textit{safe access bit} set. %
 The delay by instructions that do not have the \emph{safe access bit} set has a high impact on the performance since the speculation is correct in the majority of the cases, i.\,e., they commit later compared to the baseline design. %
 This leads to the higher than average performance overhead of~6.5\,\%. %

\subsection{Mutually Distrusting Sandboxes} 
\label{sec:mutually-distrusting}
This subsection showcases how two state-of-the-art sandboxing frameworks can easily be extended with Okapi to enforce additional in-process isolation for the speculative domain. %

\subsubsection{WebAssembly}
\label{sec:wasmtime}
We modified the Wasmtime Web\-Assembly runtime~\cite{2017-wasmtime} as shown in~Fig.~\ref{fig:Okapi-wasm-tmp}. %
In this use case, the trusted runtime executes a set of untrusted workloads (jobs) in a round robin schedule. %
Whenever a job runs out of so called \emph{fuel}, it hands back the control to the runtime (black arrows). %
Every time the runtime calls or suspends a job, it executes an \emph{OkapiReset} instruction (green arrows). %
This ensures that the newly assigned job has no speculative access rights. %
As a result, a malicious job can neither transiently access another job's data nor the data belonging to the runtime (red arrows). %
\par
In this experiment we use the \textit{Shootout} workloads from the Sightglass Benchmark Suite~\cite{2019-sightglass} and reduce the iterations performed by the more computation-intensive workloads to keep the simulation effort feasible. %
We select the benchmarks with a CPU time greater than one second to capture the most realistic workloads with a representative amount of switches and \emph{OkapiReset} instructions.
We set up the Wasmtime environment to execute the workloads interleaved. %
We run the benchmarks with a varying amount of fuel  on our Okapi CPU as well as on an unmodified baseline CPU for comparison. %
\par  
The results are depicted in~Fig.~\ref{fig:okapi_shootout}. On average, Okapi incurs an overhead ranging from 5.52\,\% for ten  thousand fuel to only 2.34\,\% for one million fuel to mitigate breakout attacks between the different tenants.
The experiment shows that running the same workloads with more fuel, i.e., fewer trust domain switches is beneficial for the Okapi architecture.

\begin{figure}[t]
    \centering
    \includegraphics[width=.6\linewidth]{fig/webAssembly.png}
    \vspace{-1em}
    \Description{
        The figure visualizes the switch between two WebAssembly Runtime jobs. Executing the \textit{OkapiReset} instruction whenever switching from the WebAssembly runtime to a job effectively blocks Spectre attack vectors between running jobs as well as Spectre attack vectors that target the WebAssembly runtime from one of the jobs.
    }
    \caption{
        WebAssembly runtime environment extended with \textit{OkapiReset} instructions when starting/continuing execution of a job (green). This mitigates the Spectre attack possibilities (red) between the trust domains (cyan) and the runtime.
    }
    \label{fig:Okapi-wasm-tmp}
\end{figure}

\begin{figure}[t]
    \centering
    \includegraphics[width=.9\linewidth]{fig/wasmtimeShootout.png}
    \vspace{-1em}
    \Description{
        The figure shows the Shootout performance results of Okapi. The workloads are run interleaved with a varying amount of fuel. The results show that Okapi incurs about 2.34\,\% overhead in the geometric mean for one million fuel and 5.52\,\% for ten thousand fuel.
    }
    \caption{
        Sightglass Shootout
    }
    \label{fig:okapi_shootout}
\end{figure}

\subsubsection{ERIM}
\label{sec:erim}
In a second case study, we investigate the security synergies of Okapi with the ERIM~\cite{2019-Vahldiek-OberwangerElnikety.etal} sandboxing library. %
Okapi prevents speculative execution of code located on different memory pages by its hardware and can therefore extend ERIM with execution prevention in case of speculatively misguided control flows, as, e.g., in a Spectre-BTB or Spectre-RSB attack. %
\par
Additionally, we instrument the \emph{erim\_switch\_to\_untrusted} macro with 
our newly introduced
\emph{OkapiReset} instruction, as depicted in~Fig.~\ref{fig:erim}. %
The \emph{erim\_switch\_to\_trusted} macro is adapted analogously.
Note that no other modification of the framework is needed to extend ERIM to Okapi.
While the memory protection key prevents the cross-domain access, resetting the \emph{safe access bits} restricts the speculatively accessible memory space. %
This ensures that previously accessed memory regions accessible by the current set of memory keys cannot be leaked speculatively within an ERIM domain. %
\par
In our experiment we run a bubblesort algorithm for arrays of 64 and 4096 bytes.
Adding \emph{OkapiReset} to the ERIM domain switches does not affect the runtime of the algorithm.

\begin{figure}[t]
    \begin{lstlisting}[language=C, gobble=13]
             #define erim_switch_to_untrusted 
             do { 
                ERIM_SWITCH_TO_UNTRUSTED_STACK;   
                __wrpkrucheck(ERIM_UNTRUSTED_PKRU);
                ERIM_INCR_CNT(1);
                asm volatile("OkapiReset");
             } while(0)
    \end{lstlisting}
    \vspace{-1em}
    \Description{
        The listing shows the code that is being executed when ERIM switches to the untrusted domain. Adding Okapis protection guarantees is as simple as adding the \textit{OkapiReset} instruction at the end of the code.
    }
    \caption{
        Okapi enhanced ERIM switch to the untrusted domain
    }
    \vspace{-1em}
    \label{fig:erim}
\end{figure}

\subsection{Vaulting}
\label{sec:vaulting}

The previous subsections consider security against breakout attacks.
They show that Okapi (i)~performs well for a representative set of workloads (cf. 7.1), (ii)~introduces only minor performance overhead when applied in realistic in-process isolation scenarios (cf. 7.2.1), and (iii)~requires only minor software modifications when combined with memory protection keys for in-process isolation (cf. 7.2.2).
\par
This subsection addresses the remaining case of poisoning attacks. 
We benchmark the Libsodium~\cite{2023-libsodium} constant-time cryptographic library which already comes with a page-aligned memory allocator. %
This feature facilitates achieving poisoning security (cf.~Sec.~\ref{sec:software}). %
\par
In this experiment, we compare three different implementations of security-sensitive code sections~$p_H$ which we refer to as \emph{atomic}~$p_H$, \emph{OkapiLoad}~$p_H$, and \emph{function-sized}~$p_H'$. %
For \emph{atomic}~$p_H$ we instrument the hashing, secret and public key routines within the cryptographic library to execute an \emph{OkapiReset} instruction after every load instruction targeting secret data to prevent poisoning attacks from leaking secret data. %
\emph{OkapiLoad}~$p_H$ achieves the same level of security by replacing every load instruction to secret data with an \emph{OkapiLoad} instruction. %
This instrumentation allows all sections in~$p_L$ to speculate freely on data accessed by their predecessors. %
Lastly, \emph{function-sized}~$p_H'$ adapts the same routines to reset the permissions on function returns. %
This instrumentation approximates the security-critical parts of the program to the set of sections~$p_H'$ consisting of the code page(s) of entire functions. %
While this measure may not guarantee full security, it drastically reduces the attack surface as well as the potential gadget space in proportion to the region size (cf.~Sec.~\ref{sec:attack-surface}). %
This can be automated w.\,r.\,t.\ to a predefined granularity. %
To this end, we modified the LLVM compiler framework~\cite{web_llvm}. %
LLVM's intermediate representation allows for an easy adaptation of the chosen granularity. %
\par
Tab.~\ref{tab:sodium} in Appendix~\ref{sec:appendix_experiments} summarizes the number of Okapi instructions executed by the different instrumentations. %
\par
We performed 100 runs of different routines in Libsodium and normalized the median 
 runtime to the unmodified version on an unmodified CPU. %
The finest possible granularity of~$p_H$ implemented by \emph{OkapiReset} incurs an average performance overhead of 61.72\,\% to achieve \textit{weak speculative non-interference}. %
While maintaining the same level of security this overhead can be reduced to~59.78\,\% by using \emph{OkapiLoad}. %
This is a clear improvement over software-based mitigations~\cite{2018-Intel-IBPB,2018-Intel-STIBP,2018-Intel-IBRS,2018-Turner} currently rolled out in industry. %

Resorting to the more coarse-grained \emph{function-sized}~$p_H'$ significantly reduces the incurred overhead even further to~11.60\,\%. %
\par
For more detail on the individual overheads refer to Fig.~\ref{fig:Sodium} in Appendix~\ref{sec:appendix_experiments}.
\par

Note that the Okapi security features against poisoning complement the security level provided by the constant-time library. %
Without Okapi, a poisoning attack can redirect the control flow such that the vault leaks the secret despite constant-time measures being in place (cf.~\cite{DBLP:conf/asplos/YavarzadehACGGK24}). %

\section{Hardware Overhead}
\label{sec:HW-overhead}
We implemented the Okapi architecture according to Sec.~\ref{sec:architecture} in the open-source out-of-order processor BOOM~\cite{2020-ZhaoKorpan.etal}. %
Afterwards, we performed synthesis and place \& route in Synopsys DesignCompiler and IC Compiler II for both the original design and our OkapiBOOM (both in \textit{medium configuration}) at 500 MHz on a commercial 12 nm FinFET process from GlobalFoundries. %
Okapi does not impact the critical path of the chip\, allowing us to achieve the same maximum clock frequency as the original BOOM core. %
Compared to the baseline design, Okapi increases cell area by only 0.6\,\% while reducing the power consumption by 1.53\,\%. %
For additional details, refer to Appendix~\ref{sec:appendix_synthesis}. %
This confirms our claim that the Okapi architecture incurs a negligible hardware overhead, in particular, when compared to DIFT (cf.~Sec.~\ref{sec:related-work}).   %

\section{Security Evaluation}
\label{sec:security-evaluation}

\subsection{Formal Hardware Verification}
\label{sec:formal}
In Appendix~\ref{sec:appendix_attack} we apply a standard Spectre attack to the gem5 o3 CPU with and without Okapi protection. %
Mounting a single example attack and showing it to be blocked by Okapi demonstrates the Okapi mechanisms but does not provide full confidence in the claimed security guarantees. %
Therefore, we also conducted a formal analysis to provide an exhaustive security proof for the Okapi architectural concept. %

We base our verification on the methodology of \emph{Unique Program Execution Checking} (UPEC), as presented in~\cite{2023-FadihehWezel.etal}, which formally establishes \emph{weak speculative non-interference} for a design under verification (DUV). %
Our DUV is BOOM extended by the described Okapi architecture.\footnote{Available on GitHub: \url{https://github.com/RPTU-EIS/OkapiBOOM}} %
We adapt UPEC to sandboxing (cf.~Sec.~\ref{sec:threat-model}) by constraining the UPEC property~\cite{2023-FadihehWezel.etal} such that the program in its sequential semantics never accesses high memory locations. %
\par
The Okapi-adapted UPEC property holds after a proof runtime of 41 minutes. %
This guarantees breakout security for sandboxes of custom size, under the assumption that the software executes an \emph{OkapiReset} instruction on sandbox exits and that it separates confidential and public data pages.
These assumptions can be easily checked for a given software. %
As the original BOOM design has been proven susceptible to Meltdown~\cite{2023-JauchWezel.etal}, our formal verification also showcases the inherent mitigation of Meltdown by Okapi. %
\par
In addition, our formally verified hardware guarantees that any program~$p$ is poisoning-secure if the software assumptions of Sec.~\ref{sec:software} are fulfilled. %
This can be proven for a given software by checking that the program sections~$p_H$ do not add the high memory locations to the trust domain. %

\subsection{Attack Surface Analysis}
\label{sec:attack-surface}
In this section, we discuss the reduction of the attack surface for Okapi in its least-overhead version, i.e., when only implementing its HW measures. %
While this version provides formally proven breakout security at thread level, the reduction of the attack surface for poisoning cannot be assessed formally. %
However, a quantitative analysis provides valuable insights. %

As explained in Sec.~\ref{sec:breakout}, the number of speculatively accessible pages of a trust domain increases over time. %
However, the empirical results for SPEC CPU2017 shown in Fig.~\ref{fig:spec-cx-switch-pages} demonstrate that programs access only 5.69\,\% of their mapped code pages and~31.27\,\% of their mapped data pages between two context switches\footnote{Every context switch involves a privilege switch}, on average. %
It is only this set of data pages visited sequentially between context switches that can be accessed speculatively by any Spectre gadget in the mapped address space. 
As Okapi without any software measures prevents the execution of gadgets following transient jumps to arbitrary code pages, this leads to a drastic reduction of the attack surface for poisoning since attackers could only exploit gadgets on the 5.69\,\% of the code pages that are reached by the program flow non-speculatively. %

If developers of security-critical applications wish to remove also this last gap, they only need to care about Spectre gadgets in the immediate vicinity of locations (i.\,e., the pages) where the code works with secret data. %
Spectre gadgets in other locations are automatically suppressed by Okapi without any further action needed by the developers or the compiler. %
\par
An interesting observation we make is that, even if an attacker could successfully access confidential data, the cache side channel can only be exploited if the address used to encode the data also belongs to the trust domain. %
Hence, even without any software support, the Okapi hardware alone (besides guaranteeing breakout security) drastically increases the effort necessary for mounting a successful poisoning attack. %

In conclusion, we believe that the hardware-only version of Okapi can be an attractive security solution in many practical settings. %
This and other sweet spots in Okapi's trade-offs are summarized in the following section. %

\section{Sweet Spots in Okapi's trade-offs}
\label{sec:sweet-spots}
As presented in previous sections, Okapi allows for several trade-offs between the level of protection and associated cost. %
The range and characteristics of these trade-offs, unfortunately, cannot be easily formalized and simply mapped to formal notions of security. %
Instead, however, we have demonstrated in our experiments that different Okapi trade-offs match particularly well with certain practical settings. %
In the following, we point to such "sweet spots" where Okapi, in practice, can help to achieve a high level of protection at exceptionally low cost: %
\begin{itemize}
\item Okapi only introduces 
an area overhead of 0.6\,\% after synthesis and place \& route.
For use cases without security requirements, the Okapi features can be switched off to avoid all performance overhead. %
\item By its hardware measures alone, Okapi provides secure speculation for the sandboxing threat model, as formally defined in ~\cite{2021-GuarnieriKoepf.etal}, at thread level. %
\par
In addition, the \textit{OkapiReset} instruction can be used to secure software sandboxes at a finer granularity. %
Our experiments in Sec.~\ref{sec:wasmtime} show a performance overhead of only 1.68\,\% to eliminate the possibility for an untrusted Wasmtime workload to launch a breakout attack, even on workloads sharing the same thread. %
\item As explained in the previous section, the above low-overhead measures do not only protect against sandboxing but, on top, provide significantly stronger security also for poisoning attacks. %
In addition, for systems without explicit sandboxing in software and for legacy code, this measure eliminates universal read gadgets and, more importantly, protects the kernel from being misused as a confused deputy. %
\item There will remain special (arguably rare) situations where the remaining gadget space within one code page is considered insufficiently secure and the user wishes to achieve address-level \textit{weak speculative non-interference}~\cite{2021-GuarnieriKoepf.etal}. %
In this case, the \textit{OkapiLoad} instruction fully closes this gap. %
The expected performance overhead may grow up to~60.0\,\%, which is fully competitive with current industrial solutions offering this level of security. %
However, such a high level of security is normally not necessary for the full application. %
Okapi allows for limiting its expensive measures to only those code sections where they are really needed, thus, in practice, avoiding excessive overhead over the full software. %
\item 
Okapi enforces (in-)process isolation for the speculative domain and thereby complements designs that already implement other dedicated isolation mechanisms. %
Intel MPK~\cite{2024-IntelMPK}, for example, guards processes with additional memory protection keys, but the same keys are available to all threads within a process. 
This gap is avoided in Okapi by additionally providing speculative separation between threads. %
Other work~\cite{2020-SchrammelWeiser.etal} enables memory protection keys to also implement in-process isolation, but lack execution prevention.
An attacker may still transiently divert the control flow to create a leakage.
Such poisoning attacks are mitigated in Okapi by the mechanisms described in~\refsec{poisoning}.
\end{itemize}


\section{Related Work}
\label{sec:related-work}
SafeBet~\cite{2023-GreenNelson.etal} exploits temporal locality based on similar observations as in Okapi. %
However, the SafeBet mitigation scheme introduces a cache-like hardware structure to keep track of which code regions have already accessed which memory regions legally. %
The performance overhead incurred by SafeBet depends on the chosen granularity of the monitored code regions, and ranges from 6\,\% (at~1\,GB granularity) to over 270\,\% (at address level). %
Also, the hardware overhead is significantly higher than that of Okapi, due to the additional 8.3\,KB high-speed cache needed for access mapping. %

SpecBox~\cite{2023-TangWu.etal} mitigates cache-based TES attacks by hiding side effects of speculative loads within the cache at cache-line granularity with minimal performance overhead.
Okapi, in comparison, protects against all TES attacks (including other, non-cache-based leakage channels, e.g., Spectre-STC~\cite{2020-FadihehMueller.etal}, and Speculative Interference Attacks~\cite{2021-BehniaSahu.etal}) by already preventing the speculative access to protected data.
Okapi protects data at a coarser granularity (page-level) which is well suited for sandboxing scenarios, such as FaaS, and therefore comes with less implementation overhead.

Hardware-only mitigations based on DIFT~\cite{2020-YuMantri.etal, 2019-YuYan.etal, 2023-JauchWezel.etal, 2021-ChoudharyYu.etal, 2021-LoughlinNeal.etal} protect data at rest with a performance overhead ranging from 10.5\,\%~\cite{2020-YuMantri.etal} to~36.0\,\%~\cite{2023-JauchWezel.etal}.
Additional hardware protection for data in transit incurs a performance overhead of up to~45.0\,\%~\cite{2021-ChoudharyYu.etal}. %
While these techniques deliver security guarantees at address-level granularity, they come at the price of higher hardware and performance overheads compared to Okapi. %
Most importantly, they significantly increase the design complexity of the microarchitecture, particularly across the memory hierarchy, which impedes their adoption. %

A second class of hardware-based mitigations utilizes manual annotations by the programmer to inform the hardware about what data to protect.
These approaches range from mitigations for Spectre-PHT~\cite{2019-FustosFarshchi.etal} to schemes that protect all annotated data in transit against speculative leakage~\cite{2020-SchwarzLipp.etal, 2023-DanielBognar2.etal}.
While SpectreGuard~\cite{2019-FustosFarshchi.etal} offers less protection than Okapi, ConTeXT~\cite{2020-SchwarzLipp.etal} inherits the hardware overhead of DIFT approaches and incurs a performance overhead of up to 71.14\,\% in security-critical applications.
ProSpeCT~\cite{2023-DanielBognar2.etal} only applied a simplified version of its architecture to custom benchmarks and achieved a performance overhead of 10.0\,\% to~40.0\,\%.

Perspective~\cite{2024-KimRudo.etal} mitigates \emph{active} as well as \emph{passive} attacks on the OS kernel. 
\emph{Active} attacks in Perspective's taxonomy  can be compared to Okapi's breakout attacks while the \emph{passive} attacks can be mapped to poisoning attacks.
Perspective achieves similar security as Okapi but its scope is limited to the kernel while Okapi can mitigate attacks on and by smaller sandboxes even if they share the same process.
Additionally, Perspective requires modifications to the kernel and a static or dynamic analysis of the system calls.
\par
Okapi effectively enforces dynamic sandboxing for speculative data accesses and therefore shares this objective with conventional sandboxing approaches and (in-)process isolation techniques that also mitigate certain Spectre attacks~\cite{2021-NarayanDisselkoen.etal,2020-SchrammelWeiser.etal,2023-NarayanGarfinkel.etal,2019-BourgeatLebedev.etal}. %
However, these mitigations pursue a different security target and introduce a much higher overhead in implementation and performance. %
In addition, while most of these techniques can thwart Spectre-PHT attacks, many of them leave an open attack surface to Spectre-BTB. 
\par
Retpolines~\cite{2018-Turner} are a software construct to prevent branch target injection. They pursue a similar goal as Okapi's hardware mechanism to restrict Spectre attacks after speculative instruction page crossings.
Since the initial introduction there has been effort to improve on the high performance overhead introduced by retpolines.
In~\cite{2021-DutaGiuffrida.etal} the authors reduce the initial overhead from 149\,\% to 10.6\,\%.
Retpolines require recompilation of the software that needs to be protected and provides incomplete protection on some CPUs~\cite{2022-WiknerRazavi}.

Further, Intel rolled out (enhanced) Indirect Branch Restricted Speculation (IBRS)~\cite{2018-Intel-IBRS}, Single Thread Indirect Branch Predictors (STIBP)~\cite{2018-Intel-STIBP} as well as Indirect Branch Predictor Barriers (IBPB)~\cite{2018-Intel-IBPB} in response to Spectre-BTB.
Both, IBRS and STIBP have been shown to still be vulnerable to branch target injection attacks~\cite{2024-LiYavarzadeh.etal}.
While IBPB is the most strict countermeasure, it introduces a high performance overhead when activated.
Hence, it is not well suited for Function as a Service (FaaS) scenarios that require frequent security domain transitions~\cite{2024-LiYavarzadeh.etal}.
\par
In 2019, Intel introduced a control-flow enforcement technology (CET)~\cite{2019-ShanbhogueGupta.etal} that restricts the possible control flow after an indirect branch to legal jump targets. While this protects software from a control flow redirection to arbitrary targets, the set of possible targets still leaves space for poisoning attacks. 
\par
Serberus~\cite{2023-MosierNemanti.etal} presents a methodology to harden constant-time code against Spectre attacks in software. 
The authors report overheads ranging from 0.9\,\% to 196.2\,\% with a standard deviation of around 80\,\% on a selection of five cryptographic primitives.
It depends on existing control flow enforcement technologies and LLVM compiler features to provide a software-based protection of static constant-time code for the constant-time threat model~\cite{2021-GuarnieriKoepf.etal}.
While the results are promising we consider Serberus as an orthogonal approach.


\section{Conclusion}
\label{sec:conclusion}
In this paper, we present Okapi, a novel hardware/software cross-layer architecture as an effective defense against breakout and poisoning attacks in sandboxed environments.
\par
At its core, Okapi prevents these attacks by delaying speculative accesses to data pages which have not yet been accessed from within the currently running sandbox.
Okapi provides the possibility to customize the granularity of sandboxes, thereby covering a range of possible applications.
The Okapi hardware architecture is easy to implement and can be seamlessly integrated into existing designs. 
In particular, breakout security, being highly relevant in applications of edge computing, is achieved with an exceptionally low hardware and performance overhead.
This bears promise for Okapi's adoption in industrial practice, possibly replacing costly software patches.  

An interesting direction for further research could be the combination of the Okapi hardware mechanisms with orthogonal software protection mechanisms, e.\,g., Serberus~\cite{2023-MosierNemanti.etal} for further performance improvements.



\newcommand{\spp}{\grantsponsor{spp}{DFG SPP Nano Security}{https://www.dfg.de/en/news/news-topics/announcements-proposals/2023/info-wissenschaft-23-11}}
\newcommand{\haspro}{\grantnum{spp}{HaSPro}}

\newcommand{\bsi}{\grantsponsor{bsi}{BSI}{https://www.bsi.bund.de}}
\newcommand{\silkostu}{\grantnum{bsi}{SILKOSTU}}

\newcommand{\intel}{\grantsponsor{intel}{Intel Corp.}{https://www.intel.de/}}
\newcommand{\scap}{\grantnum{intel}{Scalable Assurance Program}}

\newcommand{\cyberagentur}{\grantsponsor{cyberagentur}{Agentur für Innovation in Cybersicherheit GmbH}{https://www.cyberagentur.de/}}
\newcommand{\evit}{\grantnum{cyberagentur}{PROTECT (EVIT program)}}

\begin{acks}
    We are grateful to all our reviewers and the shepherd for substantial feedback and help with revising the paper.
We further thank Jan Lappas and Norbert Wehn for their valuable support. 
    The reported research was partly supported by the
    \spp{} under grant \haspro{}, by the \cyberagentur{} under grant \evit{},
    by the
    \bsi{} under grant \silkostu{}
    and by the
    \intel{} \scap{}.
\end{acks}
\newpage

%
%
%

%
%
%

%
%

%
%
\bibliographystyle{ACM-Reference-Format}
\bibliography{refs3}
%
%
%
%
%


\appendix
\section{Additional experimental Results}
\label{sec:appendix_experiments}

\reffig{ROBLoads} shows the number of safe and unsafe loads that are dispatched during the SPEC CPU2017 benchmarks. It is visible that the vast majority of loads entering the ROB are unsafe loads of which on average 85.5\,\% are committed eventually.
\par
\reffig{OkapiLoads} shows the status of speculative load instructions when the LSQ tries to execute their address translation for the first time. On average, around 14.3\,\% of the speculative load instructions are blocked from going to the TLB due to our countermeasure to prevent the use of Spectre-BTB gadgets. Out of the remaining load instructions that are passed to the TLB speculatively, 98.4\,\% have their safe access bit set.
\par
Tab.~\ref{tab:sodium} provides the exact number of Okapi instructions added by the different instrumentations. \emph{atomic~$p_H$} adds an \emph{OkapiReset} after every load instruction, while \emph{OkapiLoad~$p_H$} replaces every load instruction by an \emph{OkapiLoad}.
Thus, both instrumentations have the same amount of special instructions. However, \emph{atomic~$p_H$} has a larger binary size because it adds instructions instead of replacing them.
The third option, \emph{function-sized~$p_H'$}, has the least amount of added instructions. 
\par
\reffig{Sodium} summarizes the runtime overhead of the median execution times of 100 runs of the different routines available in the Libsodium library normalized to the unmodified version on an unmodified CPU. As can be seen, the finest possible granularity of~$p_H$ implemented by OkapiReset introduces an average performance overhead of 168.43\,\%. While maintaining the same level of security this overhead can be reduced to~59.78\,\% by using OkapiLoad. Resorting to the more coarse-grained function-granular choice of $p_H'$ significantly reduces the incurred overhead even further down to~11.60\,\%

\begin{figure}[t]
    \centering
    \includegraphics[width=.95\linewidth]{fig/ROBLoads.png}
    \vspace{-1em}
    \Description{
        The figure shows the number of safe and unsafe loads that are dispatched during the SPEC CPU2017 benchmarks. It is visible that the vast majority of loads entering the ROB are unsafe loads of which on average 85.5\,\% are committed eventually.
    }
    \caption{
        Okapi: dispatched load instructions
    }
    \label{fig:ROBLoads}
    \vspace{-0.5em}
\end{figure}

\begin{figure}[t]
    \centering
    \includegraphics[width=.95\linewidth]{fig/OkapiLoads.png}
    \vspace{-1em}
    \Description{
        The figure shows the status of speculative load instructions when the LSQ first tries to execute their address translation. On average, around 14.3\,\% of the speculative load instructions are blocked from going to the TLB due to our countermeasure to prevent the use of Spectre-BTB gadgets. Out of the remaining load instructions that are passed to the TLB speculatively, 98.4\,\% have their safe access bit set.
    }
    \caption{
        Okapi: first-time issued load instructions
    }
    \label{fig:OkapiLoads}
    \vspace{-0.5em}
\end{figure}

\begin{table}[!tpb]
    \footnotesize
    \centering
    \caption{Libsodium Okapi Instructions}
    \label{tab:sodium}
    \begin{tabular}{lrrr}
        \toprule
        Config               & atomic~$p_H$  &  OkapiLoad~$p_H$ & function-sized $p_H'$ \\           
                             & \#OkapiReset  & \#OkapiLoad      & \#OkapiReset          \\
        \midrule
        Hashing              &               &                  &                       \\
        \cmidrule(rl){1-1}
        sha256\_64B          & 633           & 633              & 8                     \\
        sha256\_8KB          & 40,384        & 40,384           & 135                   \\
        \midrule
        Secret crypto        &               &                 &                       \\
        \cmidrule(rl){1-1}
        salsa20\_stream      & 186           & 186              & 6                     \\
        salsa20\_stream\_xor & 194           & 194              & 6                     \\
        secretbox\_easy      & 670         & 670              & 27                    \\
        secretbox\_open      & 808         & 808              & 29                    \\
        aead\_ietf\_encrypt  & 337           & 337              & 35                    \\
        aead\_ietf\_decrypt  & 337           & 337              & 37                    \\
        \midrule
        Public crypto        &               &                  &                       \\
        \cmidrule(rl){1-1}
        box\_easy            & 41,034       & 41,034           & 1,324                  \\
        box\_open            & 41,074       & 41,074           & 1,313                  \\
        \midrule
        Mean                 & 12,576        & 12,576           & 292                   \\ 
        \bottomrule
    \end{tabular}
\end{table}

\begin{figure}[t]
    \centering
    \includegraphics[width=.95\linewidth]{fig/Libsodium.png}
    \vspace{-1em}
    \Description{
        This figure summarizes the runtime overhead of the median execution times of 100 runs of the different routines available in the Libsodium library normalized to the unmodified version on an unmodified CPU. As can be seen, the finest possible granularity of~$p_H$ implemented by OkapiReset introduces an average performance overhead of 168.43\,\%. While maintaining the same level of security this overhead can be reduced to~59.78\,\% by using OkapiLoad. Resorting to the more coarse-grained function-granular choice of $p_H'$ significantly reduces the incurred overhead even further down to~11.60\,\%.
    }
    \caption{
        Libsodium overheads
    }
    \label{fig:Sodium}
    \vspace{-0.5em}
\end{figure}

\section{Spectre-PHT Sample Attack}
\label{sec:appendix_attack}

We mounted a Spectre-PHT sample attack from the Google SafeSide repository~\cite{web-safeside} in order to validate the security provided by Okapi. %
The application holds a public data array as well as a secret data array.
We modified the code such that it places the secret data on a separate page. 
After checking that the attack works as expected on the unmodified out-of-order model, the CPU model is changed to an Okapi out-of-order CPU. %
The in-process PHT attack performs legal accesses to the public data array to mistrain the predictor until it changes the offset to point to a byte in the secret string. %
The secret byte is then retrieved by a cache leakage gadget measuring the latency for the entries in an array data structure. %
Fig.~\ref{fig:timing} shows the average access time to the elements of that array over ten runs. %
The unmodified out-of-order CPU model has an average access time of 64 cycles for the 73rd element which represents the character~'I' indicating a successful attack. %
For the modified Okapi CPU, on the other hand, no element of the timing array is cached. %
This can be seen in Fig.~\ref{fig:timing} where all access latencies are above the threshold, documenting that the Okapi architecture successfully prevents the attack. %

\begin{figure}[t]
    \centering
    \includegraphics[width=.95\linewidth]{fig/spectre_pht.png}
    \Description{The figure shows the average access latency for the timing array in the Spectre-PHT attack over 10 runs. While the unmodified o3-CPU shows leakage, the Okapi-CPU does not.}
    \caption{Average access latency for the timing array in the Spectre-PHT attack over 10 runs.}
    \label{fig:timing}
\end{figure}

\section{Synthesis and Place \& Route}
\label{sec:appendix_synthesis}

Synthesis of the BOOMv3 core~\cite{2020-ZhaoKorpan.etal} shows that the front-end dominates the critical path. %
With both instruction and data cache, as well as parts of the TLB module implemented as SRAM cells, BOOMv3 can easily reach a clock frequency of 500\,MHz after synthesis and place \& route on a commercial 12\,nm FinFET Globalfoundries process for the \textit{medium configuration} of the core. %

We used the worst-case corner as the target corner for the timing optimizations with a supply voltage of 0.72\,V slow-slow transistor corner and 125\,\textdegree C temperature. 
The employed ARM standard cell library uses only very low threshold transistors for minimum dynamic power and maximum performance.
The power results are generated in Synopsys PrimePower with typical corner settings (TT, 0.8\,V, 25\,\textdegree C).

The runtime for the synthesis was roughly 6~hours, with additional 2.5~days for place \& route and static timing analysis. %
Our results for BOOMv3 align well with the reported results for BOOMv2. %
Here, the developers report a tapeout frequency of 594~MHz with optimized place \& route setup in a 28\,nm process~\cite{2018-Celio}. %

Running the same process on our modified OkapiBOOM core shows that the additional hardware does not deteriorate the maximum clock frequency as the implemented design features are not on the critical path. %
Compared to the baseline design, Okapi adds 0.6\,\% of cell area. %
It is interesting to mention that the baseline design takes 1.5\,\% more area than OkapiBOOM after place \& route, due to a more complicated placement and wiring. %
In addition, incorporating the Okapi features reduces the power consumption by 1.5\,\%. %
A detailed comparison is listed in Tab.~\ref{tab:sythesis}. %
The small differences are all within the tolerance of the commercial tools (noise), and are not a systematic cause of the Okapi extension.

Our results clearly show that the additional hardware does not jeopardize the synthesizability of the BOOM design. %
The resulting layout of OkapiBOOM after place \& route is shown in Fig.~\ref{fig:OkapiBOOMfloorplan}. %

\begin{table}[t]
    \footnotesize
    \centering
    \caption{Comparison of Synthesis Results in 12\,nm FinFET}
    \label{tab:sythesis}
    \vspace{-1em}
    \begin{tabular}{lrrr}
        \toprule
                                       & Baseline      &  OkapiBOOM   & \\           
        \midrule
        Area*                           &               &              & \\
        \cmidrule(rl){1-1}
        Combinational Area ($GE$) & 859,219    & 891,426   & +3.8\,\%   \\
        Register Area ($GE$)      & 1,188,868  & 1,172,435 & -1.4\,\%   \\
        Macro Area ($GE$)         & 554,294    & 554,294   & 0.0\,\% \\
        Total Cell Area ($GE$)    & 2,602,465  & 2,618,088 & +0.6\,\%   \\
        Total Area ($GE$)         & 4,730,557  & 4,659,064 & -1.5\,\%   \\
        \midrule
        Power**                        &               &                  \\
        \cmidrule(rl){1-1}
        Dynamic Power*** ($mW$)        & 246.7         & 242.6        & -0.6\,\%   \\
        Leakage Power ($mW$)           & 0.008         & 0.008        & 0.0\,\% \\
        \midrule
        \multicolumn{4}{l}{\footnotesize{*  Gate Equivalents (GE), normalized to single NAND2\_X1 size}} \\
        \multicolumn{4}{l}{\footnotesize{** at 500\,MHz, TT corner (25\,\textdegree C, VDD = 0.8\,V)}} \\
        \multicolumn{4}{l}{\footnotesize{***total power = switching power + internal power}}
    \end{tabular}
\end{table}

\begin{figure}[t]
  \includegraphics[width=.95\linewidth]{fig/Layout.png}
  \vspace{-1em}
  \Description{
    Floorplan of the OkapiBOOM after place \& route. It can be seen that the additional Okapi hardware does not jeopardize the synthesizability of the BOOM design.
  }
  \caption{
    Floorplan of OkapiBOOM after place \& route
  }
  \label{fig:OkapiBOOMfloorplan}
\end{figure}

\end{document}